\documentclass[journal]{IEEEtran}
\IEEEoverridecommandlockouts
\usepackage[T1]{fontenc}
\usepackage[utf8x]{inputenc}
\usepackage{authblk}
\usepackage{cite}
\usepackage{amsmath,amssymb,amsfonts}
\usepackage{algorithmic}
\usepackage{graphicx}
\usepackage{textcomp}
\usepackage{xcolor}
\def\BibTeX{{\rm B\kern-.05em{\sc i\kern-.025em b}\kern-.08em
    T\kern-.1667em\lower.7ex\hbox{E}\kern-.125emX}}

\usepackage{hyperref}
\usepackage{tabularx}
\newcolumntype{R}{>{\raggedleft\arraybackslash}X}

\usepackage{balance}
\usepackage{multirow}
\usepackage{booktabs}
\usepackage{algorithm}
\usepackage{algorithmic}

\usepackage{xspace}
\newcommand\phoebe{{\sc Phoebe}\xspace}
\newcommand\phoebebf{{\sc \textbf{Phoebe}}\xspace}

\usepackage{amsthm}
\theoremstyle{definition}

\usepackage{array, etoolbox}
\newcounter{rowcount}
\usepackage{fancyhdr}
\AtBeginDocument{%
}

\newcommand{\commentmartin}[1]{\textcolor{blue}{martin: #1}}\newcommand\mm\commentmartin

\begin{document}

\title{Maximizing Error Injection Realism for Chaos Engineering with System Calls}

\author[1]{Long Zhang}
\author[2]{Brice Morin}
\author[1]{Benoit Baudry}
\author[1]{Martin Monperrus}
\affil[1]{KTH Royal Institute of Technology, Sweden}
\affil[2]{Tellu, Norway\vspace{0.3cm}}


\maketitle
\thispagestyle{fancy}

\begin{abstract}
In this paper, we present a novel fault injection framework for system call invocation errors,  called \phoebebf. \phoebebf is unique as follows; 
First, \phoebebf enables developers to have full observability of system call invocations. Second, \phoebebf generates error models that are realistic in the sense that they mimic errors that naturally happen in production. Third, \phoebebf is able to automatically conduct experiments to systematically assess the reliability of applications with respect to system call invocation errors in production. We evaluate the effectiveness and runtime overhead of \phoebebf on two real-world applications in a production environment for a single software stack: Java. The results show that \phoebebf successfully generates realistic error models and is able to detect important reliability weaknesses with respect to system call invocation errors. To our knowledge, this novel concept of ``realistic error injection'', which consists of grounding fault injection on production errors, has never been studied before.
\end{abstract}


\section{Introduction}

In cloud-based software systems, one cannot fully control the production execution environment and many unexpected events keep happening: hardware issues, network fluctuations, and unanticipated user behavior \cite{Leesatapornwongsa:The_Case_for_Drill-Ready_Cloud_Computing,xu18feast:binary-hardening}. In order to assess and improve the reliability of software systems in such a changing and imperfect environment, different kinds of techniques are being researched, in particular fault injection \cite{Natella:Assessing_Dependability_with_Software_Fault_Injection_A_Survey,Vieira2012:resilience-benchmarking}. Fault injection evaluates software reliability by actively injecting errors into the software system under study \cite{Ziade:A_Survey_on_Fault_Injection_Techniques, Laprie:issre1995:conceptsInDependability}. A recent trend in fault injection consists of injecting faults in production directly \cite{allspaw2012fault,Basiri2016,Zhang:chaosmachine}, this is known in the industry as 
``chaos engineering''. In this paper, we use ``chaos engineering'' for referring to fault injection in production.

It is known that the space of all possible error injection is large \cite{Alvaro:LDFI}. In other words, it is potentially intractable to explore all possible error scenarios. When doing fault injection in production, one does not want to impact users with unrealistic errors. In this paper, we address the problem of defining a tractable error injection space, by exclusively focusing on realistic errors. Our novel definition of realistic errors is that the injected errors resemble the ones that naturally happen in production. By injecting realistic errors, we identify reliability issues that are more relevant for developers. 

In this paper, we realize this idea in the realm of system call errors. This focus is motivated by the essential role of system calls to analyze software behavior \cite{forrest:sense-of-self}, and by the significant number of invocations to system calls that naturally fail in production. We propose a novel fault injection framework called \phoebe, for doing realistic error injection at the system call level. 
To define realistic errors, \phoebe first observes the natural system call invocation errors which happen in a production system. Then it analyzes those previously observed errors to synthesize a series of realistic error injection models that systematically amplify natural errors.
The injection of such realistic errors brings valuable insights into the error handling capabilities of an application with respect to realistic system call invocation errors.
To our knowledge, the synthesis of those realistic error models is the key novelty of \phoebe. 

We evaluate \phoebe with two real-world applications: Hedwig, an email server that uses the SMTP and IMAP protocols, and TTorrent, a file downloading client based on the BitTorrent protocol. During the experiments in a production environment, \phoebe observes that $84$ million invocations to $23$ unique system calls naturally fail. Based on these errors, \phoebe synthesizes $32$ and $33$ realistic error injection models for HedWig and TTorrent respectively. The generated error models are then used to perform a series of chaos engineering experiments, which reveal important reliability shortcomings in both applications. The results of these experiments demonstrate the feasibility, applicability and added value of \phoebe for analyzing reliability against system call invocation errors.

To sum up, our contributions are the following.

\begin{itemize}

\item The concept of synthesizing realistic error injection models for system calls, based on the amplification of errors that are naturally observed in production.

\item \phoebe, a fault injection framework that implements this concept. \phoebe monitors a production system, generates realistic error models, conducts chaos engineering fault injection experiments, and outputs a reliability assessment with respect to system call invocation errors. \phoebe is publicly available for future research in this area\footnote{\url{https://github.com/KTH/royal-chaos/tree/master/phoebe}}.

\item An empirical evaluation of \phoebe with two real-world applications, an email server and a file downloading client, under a production workload. The results show that \phoebe can inject realistic errors at runtime, in production, with low overhead, in order to detect error handling weaknesses with respect to system call invocation errors.
\end{itemize}

The rest of the paper is organized as follows: \autoref{sec:background} introduces the background. \autoref{sec:design-and-implementation} and \autoref{sec:evaluation} present the design and evaluation of \phoebe. \autoref{sec:discussion} discusses the runtime overhead of \phoebe and the threats to the validity of this research work. \autoref{sec:related-work} presents the related work, and \autoref{sec:conclusion} summarizes the paper.

\section{Background}\label{sec:background}

\subsection{Observability in Software Systems}
A software system is said to be observable if it is possible to analyze the system's internal state on the basis of its external behavior \cite{MilesRuss2019ChaosObservability}. For example, by observing an HTTP $500$ response code instead of $200$, developers are able to know that there are some errors in the system when handling an HTTP request. In the context of errors, observability relates to the ability of detecting when an error naturally occurs in a software system. In this case, observability helps developers to evaluate the system's error detection and handling capabilities.

There are mainly three categories of observability techniques \cite{simonsson2019observability}:
1) logging the system's internal state. For example, using \texttt{Exception.printStackTrace()} method in Java to log stack information when an exception occurs.
2) monitoring metrics exposed by a system. For example, monitoring the memory usage of an application in order to detect memory leak issues.
and 3) tracing externally observable events. For example, tracing an HTTP request that propagates through micro-services for timeout-related bug analysis.

\subsection{Chaos Engineering}\label{sec:software-fault-injection}
Chaos engineering is a recent fault injection technique, which consists of injecting faults in production in order to verify actual reliability \cite{Basiri2016}. Different from traditional fault injection techniques, chaos engineering focuses on evaluating systems while they are deployed into production. This is necessary for large-scale software systems because they are practically impossible to set up realistically in a testing environment. Chaos engineering usually builds the fault injection model based on real events (such as server crashes). Since faults are injected in production, special techniques are employed so as to keep side effects acceptable. Chaos engineering has been extensively studied with respect to specific fault models: server crashes \cite{netflix-chaos-monkey-blog}, disk space issues \cite{gremlin-infrastructure-attack} and network fluctuations \cite{Chang:Chaos_Monkey_Increasing_SDN_Reliability_Through_Systematic_Network_Destruction}.

In chaos engineering, the injected faults are called ``perturbations'' or ``turbulences''. As chaos engineering injects perturbations directly in production, there are several key principles to follow \cite{ChaosEngineeringPrinciples}: 1) define at least one hypothesis that specifies how an injected perturbation is expected to affects the application's normal behavior, 2) the injected perturbation should be based on real events, 3) the experiments should be directly done in production environment, 4) the experiments should be run continuously and automatically, and 5) the side effects of an experiment should be minimized \cite{ChaosEngineeringPrinciples}.

\subsection{System Calls}\label{sec:system-call}

System call is a fundamental interface between an application and the kernel \cite{syscallsman}. In modern operating systems, critical resources such as hardware devices and process scheduling are usually managed by the kernel. When a user application needs to interact with a given critical resource, the corresponding system call is invoked. For example, in Linux, the \texttt{open} system call is invoked when an application needs to open or create a file. Linux defines and implements more than $300$ unique system calls, as well as over $100$ error codes to precisely report on errors upon invocation of those system calls \cite{errorno-manpage}.

As discussed in \autoref{sec:software-fault-injection}, an application may be perturbed by both hardware errors and software errors. Sometimes such errors can not be handled by the operating system on its own. Thus it propagates the error to the application, by failing a system call invocation with an error code.

\section{Design \& Implementation of \phoebe}
\label{sec:design-and-implementation}

Now we describe \phoebe, a system for defining and injecting realistic system call perturbations in software systems.

\subsection{Objective of \phoebe}\label{sec:working-example}
Applications contain a large amount of code for handling system call invocation errors \cite{saha2013hector}. Some system call invocation errors happening in production do not have an visible impact on the application’s behavior, because the resilience mechanisms embedded in the application handle them properly \cite{mitra2005robust}. However, it is not possible for developers to know whether these fault-tolerance mechanisms behave properly if these errors were to occur more frequently.

If a given type of system call invocation error rarely happens, it may take a long time to observe the corresponding abnormal behavior. Similarly, even if a type of system call error happens very frequently, it is not clear whether there is a threshold on the error rate below which the application can handle the errors and beyond which it starts to behave abnormally. The key problem addressed by \phoebe is to evaluate whether system call errors can have a severe impact on the application's behavior if the error rate is higher than usual.

In order to bring insights about how an application behaves when a system call invocation fails, \phoebe follows the principles of chaos engineering to actively inject system call invocation errors into an application running in production.

First, \phoebe provides observability on system calls. It collects system call invocations including their name, execution time and return code, with low overhead. Based on these observations, \phoebe analyzes the natural system call invocation errors and builds a model of their occurrence over time.

Secondly, \phoebe injects errors inspired from natural errors, resulting in the natural system call invocation errors happening more frequently. \phoebe's error injection is designed to be realistic as follows: 1) \phoebe selects the target system call invocation and the injected error code based on the observed natural errors, and 2) \phoebe computes the rate of error injection based on the frequency of the original errors. Done in production, this is chaos engineering for system calls (per the definition of \autoref{sec:software-fault-injection}). \phoebe provides the developers with information about the behavior of the system under system call perturbations. If the application's behavior is acceptable under fault injection, the developers increase their confidence in the application's error handling capabilities. Otherwise, with the help of \phoebe, developers learn more about how the application behaves when system calls fail, and can fix the uncovered reliability issues.

\subsection{Overview of \phoebe}

\begin{figure*}[htb]
\centering
\includegraphics[width=15.5cm]{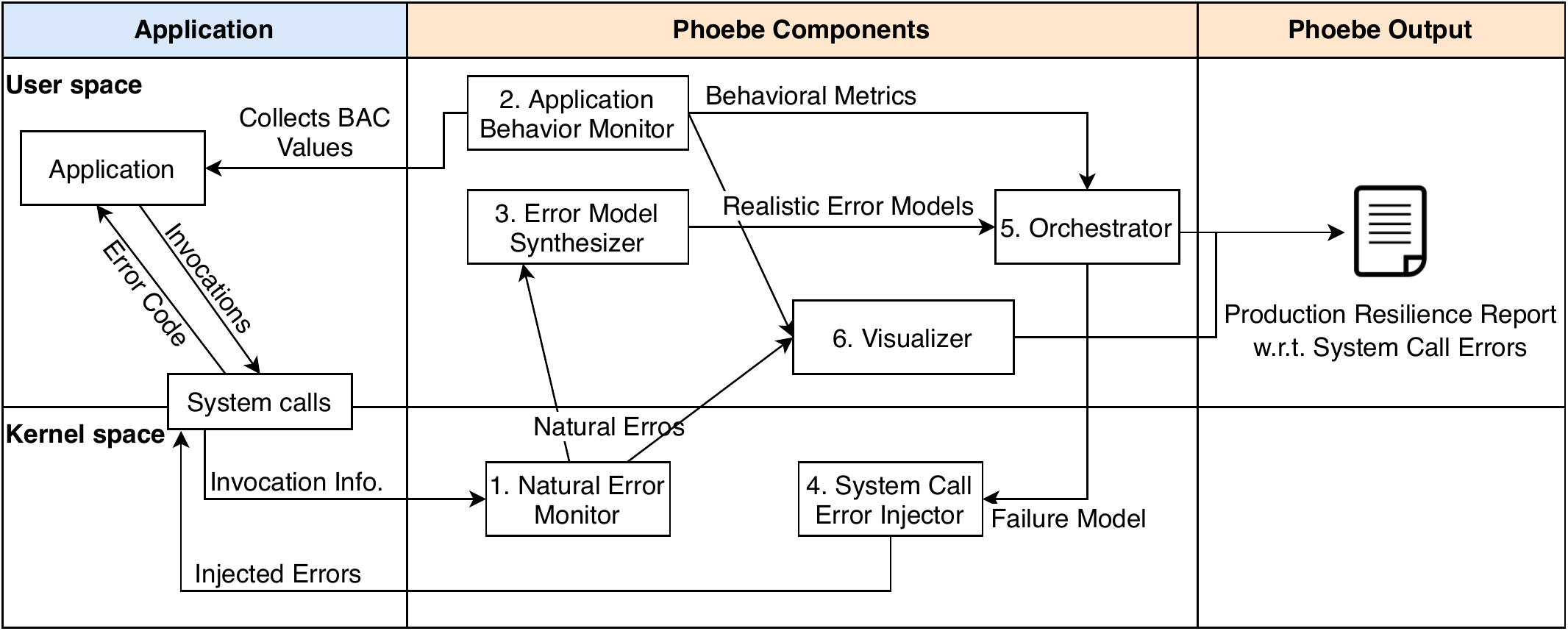}
\caption{Architecture of \phoebe: it collects system calls in production and perturbs them in a realistic way, in order to produce a valuable reliability assessment.}\label{fig:architecture}
\end{figure*}

\autoref{fig:architecture} shows the architecture of \phoebe, there are six components in \phoebe:
1) natural error monitor (see \autoref{sec:syscall-monitor}),
2) application behavior monitor (see \autoref{sec:app-monitor}),
3) realistic error injection model synthesizer (see \autoref{sec:synthesizer}),
4) system call invocation error injector (see \autoref{sec:injector}), 
5) experiment orchestrator (see \autoref{sec:orchestrator}) and 
6) metrics visualizer (see \autoref{sec:visualizer}).

\textbf{Workflow.} Briefly, those components work as follows.
The natural error monitor collects and reports system call related metrics, those metrics are inputs for the error injection model synthesizer. The synthesizer analyzes the natural system call errors and computes a realistic error injection model. 
The system call invocation error injector takes an error model as input, and conducts a set of chaos engineering experiments on the application. The experiment flow is automated by the experiment orchestrator. Finally, the metric visualizer provides a live dashboard for the developer to display the monitoring information.

\textbf{Benefits.}
\phoebe enables chaos engineering experiments on system calls, with two key benefits.
First, \phoebe improves observability with respect to system calls: with \phoebe developers are now able to observe system call invocation errors that  naturally occur  in a production environment.
Second, \phoebe performs effective chaos engineering experiments, based on error models that are generated according to realistic errors. This is an essential conceptual and technical novelty compared to  existing chaos engineering tools, which require developers to define the injected error model, regardless of its realism.

\subsection{Definitions}\label{sec:definitions}

\textbf{System Call Invocation Error}
An invocation of a system call is deemed an error if it returns an error code \cite{syscallsman}. These error codes are systemically specified and documented in the operating system under consideration, Linux in our case. In Linux, the error codes are negative values defined in a header file named \texttt{errno.h} \cite{errorno-manpage}. In this paper, an invocation to a given system call \texttt{syscall} yielding a given error code \texttt{E} is denoted by \texttt{syscall:E}.

\textbf{Monitoring Interval} A monitoring interval is a period of time along which metrics of interest are collected. The length $t$ of a monitoring interval in seconds means that \phoebe collects metrics over $t$ seconds, and reports them to an external component such as a time-series database every $t$ seconds. The smaller $t$ is, the more frequently the monitor reports the metrics.

\textbf{Error Rate}
An error rate $r$ of a system call invocation error $s:e$ during the length of monitoring interval $i$ is denoted as $r(s,e,i)$. The value of $r(s,e,i)$ is calculated as $$r(s,e,i)=\frac{number\_of\_errors(s,e,i)}{total\_number\_of\_invocations(s,i)}$$ where $number\_of\_errors(s,e,i)$ means the total number of errors $s:e$ in $i$. $total\_number\_of\_invocations(s,i)$ records the total number of invocations of system call $s$ made in $i$ no matter what a return code is.

\textbf{Realistic Error Injection Model} In this paper, an error injection model is a triple $(s,e,r)$ that states what invocation error $s:e$ is injected, with an error rate of $r$. An error model is considered as realistic if the error $s:e$ is naturally observed in a production-like running environment.

\textbf{Behavioral Assessment Criteria (BAC)} 
The normal behavior of a software system consists of functionalities that can be successfully executed by users \cite{forrest:sense-of-self}.
In this paper, we define the ``Behavioral Assessment Criteria'' as a set of application-specific metrics that capture the normal behavior and potential deviations from the perspective of users. The series of metric values observed in the absence of perturbations forms the normal behavior.
The longer we observe these metrics, the more accurately the normality of the behavior is assessed. An application's normal behavior is often described at the application level, with application specific metrics: for instance, Netflix uses metric SPS -- stream starts per second -- as their major metric for capturing the normal behavior of their video streaming system \cite{Basiri2016}.

\textbf{Chaos Engineering Experiment} In this paper, a ``chaos engineering experiment'' is a consecutive sequence of monitoring intervals during which system call invocation errors are actively injected in production according to a given error injection model. The goal of a chaos engineering experiment is to analyze to which extent the injected errors make the system deviate from its behavioral assessment criteria in production.

\subsection{Component Design}\label{sec:component-design}
\subsubsection{Natural Error Monitoring}\label{sec:syscall-monitor}
The natural error monitor captures metrics related to system calls. Given a monitoring interval length $t$, the monitor regularly captures a set of metrics including 1) the name and amount of different system call invocations during $t$, 2) the return code $e$ of system call invocations, 3) the execution time $l$ of each system call and 4) the error rate $r$ if a specific system call invocation fails.

Per our definition of a monitoring interval in~\autoref{sec:definitions}, the choice of a length $t$ for the monitoring interval results from an engineering trade-off. A smaller $t$ gives a more accurate description of metric changes thanks to a more frequent sampling. However, this brings more monitoring overhead with respect to performance and storage. For example, a small interval (e.g., 1 second) makes the monitor report metrics very frequently: lots of monitoring data is generated, which requires significant calculation and storage resources. Meanwhile, a large interval length (e.g. 1800 seconds) leads to more sparse monitoring data, which gives less accurate information about errors. Per our pilot experiment, $t=15$ seconds is a good trade-off between overhead and precision, this is \phoebe's default monitoring interval length, but it is fully configurable for end-user.

\subsubsection{Application Behavior Monitoring}
\label{sec:app-monitor}

\phoebe collects a set of behavioral assessment criteria to model the application behavior. The criteria combine general metrics, language-specific metrics and application-specific monitoring metrics which capture the normal behavior:
1) general metrics: \phoebe captures OS-level metrics such as CPU load
2) runtime-specific: \phoebe has dedicated support for Java and captures heap memory usage, garbage collection time
3) application-specific metrics: \phoebe can capture HTTP requests for certain Java libraries. The developers can define more behavior monitoring, for example, the ratio of successful database requests per second can be considered as a metric for an enterprise applications.

Both language-specific and application-specific behavior monitoring requires observability features that are specific to an execution environment. In particular, they require code instrumentation. The current prototype implementation of \phoebe supports software systems that are running on a Java Virtual Machine with support for JVM metrics and bytecode instrumentation.

\subsubsection{Realistic Error Model Synthesis}\label{sec:synthesizer}

There are many different error codes for each system call. In principle, it is possible to inject all possible error codes for all system calls. However, this is very time consuming, and some combinations never happen in the real world. In order to increase the efficiency and relevance of chaos engineering experiments, the error model synthesizer focuses on errors that occur naturally. It takes the observed natural errors as input, and generates a set of realistic error injection models for chaos engineering experiments.

Recall that an error injection model is a triple $(s,e,r)$, in which $s$ stands for the system call, $e$ means the error code of a system call invocation, and $r$ is the error rate. Given an observed system call $s$ that failed with error code $e$, the synthesizer keeps $s$ and $e$, and generates different rate values by applying an error rate amplification algorithm which is described in \autoref{algo:model-synthesis}. For each system call invocation error, we consider four different cases:

\textbf{Case 1: sporadic errors} (condition at line $5$) The natural error rate is very low: this means that it takes a very long time to observe such an error naturally. In this case, the synthesized error injection model uses a fixed error rate, large enough to inject and observe the error during the experiment. To stay realistic, the error injection rate is kept low.

For example, if the natural error rate for system call $S$ with error code $E$ is $0.000001$, it is difficult to observe this error in a normal running environment. Thus, the synthesizer amplifies it to a fixed low value, e.g., $1\%$.  This means that during a fault injection experiment, an invocation of system call $S$ has a $1\%$ likelihood to be failed with error code $E$. This amplification rate is  configurable by the developer.

\textbf{Case 2: realistic maximum for fluctuating errors} (condition at line $7$) \phoebe observes a significant difference between the maximum and minimum error rates over different monitoring intervals. In this case, the synthesizer sets the rate of the model to the observed maximum error rate. In other words, the error injection rate is realistic because it is not higher than natural error rate over a given period.

For example, system call $S$ fails with error code $E$ several times during a consecutive sequence of monitoring intervals. The minimum error rate in a monitoring interval is $0.01$ and the error rate in another interval is $0.5$. This means that system call $S$ sometimes fails once for every 100 invocations or for half of such system call invocations. The synthesizer sets the error rate to $0.5$ in the model. During a chaos engineering experiment, invocations to this $S$ always have the highest possibility of error.

\textbf{Case 3: realistic amplification of steady errors} (``else'' statement at line $9$) The maximum and minimum error rates over a consecutive sequence of monitoring intervals are close to each other, and the rate is higher than a threshold. This means that an error happens often and at a steady pace. In this case, the synthesizer multiplies the maximum original error rate by a fixed amplification factor $f$.

\autoref{tab:error-cases} summarizes the main cases of system call invocation errors and their corresponding error amplification strategies. Each row describes one error case which is presented above, including its original error rate, the corresponding amplification strategy, and the benefit for developers to inject perturbation based on an amplified error rate.

\begin{table*}[tb]
\centering
\scriptsize
\caption{When System Call Invocation Errors Happen in Production: Taxonomy ($b$ means the upper boundary for using the fixed error rate $v$, $t$ is a threshold, and $f$ stands for an amplification factor.)}\label{tab:error-cases}
\begin{tabularx}{\textwidth}{lrrX}
\toprule
\textbf{Case}& \textbf{Ori. Error Rate}& \textbf{Amplification Strategy}& \textbf{Benefits of Amplification}\\
\midrule
Sporadic errors& $r_{max} < b$& Fixed error rate $v$& Make rare errors actually happen so that enough data could be captured for behavior analysis.\\
\midrule
\multirow{2}{*}{Fluctuating errors}& $r_{max} > b$& \multirow{2}{*}{$r_{max}$}& \multirow{2}{*}{Make fluctuating errors always happen more steadily so as to observe behavioral chances.}\\
& $variance\_of(r) > t$& & \\
\midrule
\multirow{2}{*}{Steady errors}& $r_{max} > b$& \multirow{2}{*}{$r_{max}*f$}& Make the errors happen more frequently to see if there is a threshold beyond which abnormal\\
& $variance\_of(r) < t$& &behavior appearance could be observed.\\
\bottomrule
\end{tabularx}
\end{table*}

Note that in \phoebe, the minimum, mean and maximum error rates are calculated from a consecutive sequence of monitoring intervals. Considering there may be extreme cases in production that make the minimum and maximum values deviate a lot from the others, the \autoref{algo:model-synthesis} takes the 5th percentile and 95th percentile respectively as the minimum and maximum error rate.

\begin{algorithm}[tb]
\caption{Realistic Error Injection Model Synthesis}\label{algo:model-synthesis}
\begin{algorithmic}[1]
\REQUIRE ~~\\ 
System call monitoring metrics in a consecutive sequence of monitoring intervals $L$;\\
An upper boundary for using fixed error rate $b$;\\
A threshold of the variance for considering original error rates fluctuating $v$;\\
A fixed error rate for rare system calls $r$;\\
Amplification factor $f$;
\ENSURE ~~\\ 
A list of error injection models $F$;

\STATE $L' \leftarrow$ Categorize $L$ by system call $s$ and return code $e$;
\STATE $L' \leftarrow$ Calculate the minimum (5th percentile), mean and maximum (95th percentile) value per monitoring interval of each $(s,e)$'s error rate;
\STATE $F \leftarrow \emptyset$
\FOR{each system call invocation error $(s,e,r) \in L'$}
    \IF{$r_{max} < b$}
        \STATE $F \leftarrow F \cup generateErrorModel(s,e,r)$;
    \ELSIF{$\mathbb{V}(r) > v$}
        \STATE $F \leftarrow F \cup generateErrorModel(s,e,r_{max})$;
    \ELSE
        \IF{$r_{max} * f < 1$}
            \STATE $F \leftarrow F \cup generateErrorModel(s,e,r_{max} * f)$;
        \ELSE
            \STATE $F \leftarrow F \cup generateErrorModel(s,e,1)$;
        \ENDIF
    \ENDIF
\ENDFOR
\RETURN $L'$;
\end{algorithmic}
\end{algorithm}

\subsubsection{System Call Invocation Error Injection}\label{sec:injector}
As introduced in \autoref{sec:system-call}, system calls are implemented by the operating system kernel. In order to trigger a specific system call invocation error, \phoebe instruments the system call invocation in order to be able to overwrite the return value of this invocation (see \autoref{sec:implementation}). The concrete workflow includes: 1) registering the injector to the target system call's return event, 2) calculating the injection conditions before the target system call invocation returns, and 3) overriding the system call return value if all the conditions have been met. 
The system call invocation error injector supports the following triggering conditions:
\emph{Process id and process name} The injector focuses on an application's process. When the injector is registered to a system call exit event, the injector uses the process id or the process name to select all system call invocations of that process.
\emph{Error rate} The injector triggers an error according to the error rate specified in an error injection model. Before a target system call returns, the injector generates a random number $p$ ($0<p\leq1$) and compares it with the error rate $r$. This condition holds when $p$ is less than or equal to $r$.
\emph{Total number of errors} The injector injects an error within a specific absolute amount of system call invocation errors. 
\emph{Successful calls only} The injector only injects errors on invocations that are initially successful. Natural system call errors are untouched and their return code remains the same.

\subsubsection{Orchestration}\label{sec:orchestrator}
The experiment orchestrator in \phoebe is designed to conduct fault injection experiments in a fully automatic manner. The orchestrator takes an experiment configuration file format as input. The configuration file contains a set of experiments, each specified by a duration and an error injection model as defined in \autoref{sec:definitions}. For each experiment, the orchestrator attaches the error injector with its error model, capturing the application's behavior with the help of the application behavior monitor, and comparing the behavior under fault injection according to the behavioral assessment criteria. When the experiment duration has passed, the orchestrator turns off the injector and outputs the experiment result.

\subsubsection{Visualization}\label{sec:visualizer}
In order to fetch and analyze the monitoring information, developers are given a metric visualization dashboard. This dashboard displays metrics in line charts so that it is convenient to investigate the change of different system call invocations. These line charts include every system call's error rate and the number of invocations over time. For other metrics such as the number of system call invocations, developers can also directly make queries to the monitoring database via the dashboard.


\subsection{Implementation}
\label{sec:implementation}
\phoebe captures and overrides system calls with the help of the eBPF module (extended Berkeley Packet Filter) provided by the Linux kernel  \cite{bpf-helpers-manpage}. The eBPF module has several advantages for system call error injection: 1) it enables \phoebe to run programs without any modifications in the kernel, 2) the overhead is relatively low \cite{CalaveraDavid2019LinuxObservabilityWithBPF}, and 3) eBPF is a mature project that is well supported. In practice, \phoebe registers BPF programs to the \texttt{sys\_enter} and the \texttt{sys\_exit} events in category \texttt{raw\_syscalls} for system call monitoring and error injection. For injecting errors, \phoebe uses BPF's function \texttt{bpf\_override\_return} to intercept a system call invocation and change the return value into an error code.
\phoebe's monitoring infrastructure and system call invocation error injector are implemented in Python. 
All the monitoring metrics are saved into Prometheus \cite{prometheus-intro}, a time series database. The visualization component is supported by Grafana \cite{grafana-intro}, an open source visualization platform. For the sake of open research, the source code and experiment results of \phoebe are publicly-available at \url{https://github.com/KTH/royal-chaos/tree/master/phoebe}.

\section{Evaluation}\label{sec:evaluation}
This section discusses the evaluation of \phoebe, which focuses on what reliability problems can be detected by injecting system call invocation errors using a realistic error model.

\subsection{Subject Programs}
In order to evaluate \phoebe, a set of representative programs needs to be selected. The selection criteria are based on: 1) the program is a real-world project that has users (it is neither a research prototype nor specifically implemented for this evaluation), 2) the program is medium-sized so that it can be deployed using the computing resources that are available in the research lab, 3) there is a production workload or equivalent that can be used for the chaos engineering experiments, 4) the subject must be monitorable with \phoebe's behavioral monitoring component for Java (see \autoref{sec:app-monitor}). Based on those systematic criteria, we select $2$ projects for the experiments: 
1) HedWig\footnote{\url{http://hwmail.sourceforge.net/}}, an email server written in Java;
2) TTorrent\footnote{\url{https://github.com/mpetazzoni/ttorrent}}, a client for downloading files using the BitTorrent protocol, implemented in Java.

\subsection{Experiment Protocol}\label{sec:experiment-protocol}

For both case studies, we follow a 4-step protocol, described below. In the following subsections, we go into the details of how we specialize these steps, according to the application-specific production-like workload and the behavioral assessment criteria for each subject program, in order to trigger different execution paths in the program. 

First, we build a realistic error injection model by observing natural system call errors (per our definition in~\autoref{sec:definitions}). To this end, \phoebe's monitor component is attached to the program to collect system call related metrics without any error injection.

Secondly, \phoebe's error injection model synthesizer takes the monitored system call information as input. It generates a set of error injection models for chaos engineering experiments.

Thirdly, the behavioral assessment criteria of the program is set up based on the monitoring metrics. The experiment orchestrator conducts error injection experiments defined in \autoref{sec:definitions} on the program according to the generated error injection models. In order to assess the performance of \phoebe's realistic error injection with respect to a baseline, we also have a random error injection strategy for each error injection model. This random error injection strategy randomly picks any error code, with equal probability,  and the rate of error injection per system call invocation is $50\%$.

At last, the program behavior under error injection is evaluated using the behavioral assessment criteria. A report is generated to show different impacts each error injection has on the program.

In order to check that side effects remain acceptable, we also conduct a health check after each error injection experiment, see  \autoref{sec:software-fault-injection}.
If an injected error has a long-term impact on the application, even after the error injection terminates, the application is restarted to reset to a normal state. The side effect can be further minimized by using state-of-art techniques such as sandboxing techniques, which is considered out of the scope of this paper. Regarding the experiment environment, all of the experiments are done in a virtual machine which has two cores of CPU and 4 GB memory running Ubuntu 18.04.

\subsection{Experiment on HedWig}

\subsubsection{Experimental Settings for HedWig}
We use version 0.7 of HedWig for the experiments. We collect a real workload for our experiments, as follows: we create an email account, to receive emails from real-world mailing lists; we let the email server run for $90$ days. As a result, $351$ emails with different headers and bodies are collected. These emails and the observed server behavior form the experimental dataset.

In order to define the behavioral assessment criteria of HedWig server, we deploy a domain specific health checker. It executes the following workflow: 1) the checker logs into the server using one test account, 2) the checker randomly picks up an email from the dataset, and forwards it to another test account hosted on the server, 3) the checker logs into the server again using the latter account and tries to fetch the latest email, and
4) the checker compares the fetched email with the original sent email to test if the email is correctly delivered. After running the health checker for 24 hours, we collect the following behavioral assessment criteria: the percentages of sending errors (SF), fetching errors (FF), validation errors (VF), and server crashes (SC).

Considering that the HedWig server may not be able to correctly deliver emails even after error injection has stopped, we collect an additional Boolean metric called state corruption (CO) during the experiments. The CO metric is calculated by a post inspection step for each experiment: after the error injector is turned off, a randomly selected email is sent, fetched, and validated as usual to test if the server is back to working normally. If the post inspection fails, CO is true. In this case the server needs to be restarted before conducting other experiments.

\subsubsection{Experimental Results}
Now we introduce the experiment results of each step.

\paragraph{Observation of natural system call errors}
\autoref{tab:naturalErrorsOfHedwig} lists the natural errors observed by \phoebe's monitor over a period of 24 hours of email traffic, with $5760$ monitoring intervals of $15$ seconds. Each row in the table describes how many invocations to a given system call yielded a given return value. The system call invocations with a \texttt{SUCCESS} return code stand for the successful ones. The minimum (5th percentile) and maximum (95th percentile) error rates, encountered in those $5760$ monitoring intervals are also reported, as well as the mean error rates, averaged over those intervals. The last column reports on how our algorithm categorizes each entry in the table per the different cases introduced above (\autoref{sec:synthesizer}).

For example, the first row of the table shows that the error type \texttt{accept:EAGAIN} occurred $370$ times, accounting for $1.73\%$ of all the invocations to \texttt{accept}. The minimum error rate encountered over the $15$ seconds monitoring intervals for that specific error was $0.125$, the maximum rate was $0.33333$. The mean error rate was $0.19591$ and the variance of the error rates is $0.00398$. In this case, there is a significant gap between the minimum rate and the maximum rate, which indicates that in some monitoring intervals HedWig invokes \texttt{accept} system calls and most of them are successful. However, in other monitoring intervals, the percentage of failed \texttt{accept} system calls with \texttt{EAGAIN} is higher. Thus, this case meets the conditions of case 2 -- fluctuating errors -- defined in \autoref{sec:synthesizer}.

\autoref{tab:naturalErrorsOfHedwig} shows different patterns of natural system call invocation errors. Some natural errors rarely happen in 24 hours, such as \texttt{mkdir} with \texttt{ENOENT} which only fails once,  while some natural errors are frequent, for example \texttt{futex} with \texttt{ETIMEDOUT}.

\begin{table}[!tb]
\centering
\scriptsize
\caption{The Observed Natural Errors for HedWig in 24 Hours}\label{tab:naturalErrorsOfHedwig}
\begin{tabularx}{\columnwidth}{lrrR}
\toprule
\textbf{Syscall \& Err.}& \textbf{Count}& \textbf{Error Rate (min, mean, max, var.)}& \textbf{Case} \\
& & \textbf{per 15 sec period}& \\
\midrule
accept:EAG.& 370 (1.73\%)& 0.12500, 0.19591, 0.33333, 0.00398& 2\\
accept:SUC.& 21K (98.27\%)& & \\
\hline
connect:ENO.& 210 (0.62\%)& 0.07692, 0.14347, 0.25000, 0.00401& 2\\
connect:SUC.& 33.4K (99.38\%)& & \\
\hline
epoll\_ctl:ENO.& 370 (50.00\%)& 0.50000, 0.50777, 0.50000, 0.00347& 2\\
epoll\_ctl:SUC.& 370 (50.00\%)& & \\
\hline
futex:EAG.& 126 (<0.01\%)& 0.00004, 0.00005, 0.00007, <1e-5& 1\\
futex:ETI.& 73.1M (48.88\%)& 0.48304, 0.48879, 0.49017, <1e-5& 3\\
futex:SUC.& 76.5M (51.12\%)& & \\
\hline
lstat:ENO.& 2 (1.06\%)& 0.06250, 0.06250, 0.06250, <1e-5& 3\\
lstat:SUC.& 186 (98.94\%)& & \\
\hline
mkdir:ENO.& 1 (16.67\%)& 0.33333, 0.33333, 0.33333, <1e-5& 3\\
mkdir:SUC.& 5 (83.33\%)& & \\
\hline
read:EAG.& 2.06K (1.21\%)& 0.02956, 0.13149, 0.20000, 0.00312& 2\\
read:ECO.& 2 (<0.01\%)& 0.03096, 0.03689, 0.04282, 0.00004& 1\\
read:SUC.& 169K (98.79\%)& & \\
\hline
recvfrom:ECO.& 10 (<0.01\%)& 0.01060, 0.02240, 0.03125, 0.00005& 1\\
recvfrom:SUC.& 340K (100.00\%)& & \\
\hline
sendto:ECO.& 9 (<0.01\%)& 0.01059, 0.01671, 0.02548, 0.00003& 1\\
sendto:EPI.& 100 (0.03\%)& 0.00663, 0.01881, 0.04645, 0.00020& 1\\
sendto:SUC.& 357K (99.97\%)& & \\
\hline
stat:ENO.& 1.8K (0.92\%)& 0.02500, 0.05170, 0.09804, 0.00098& 3\\
stat:SUC.& 194K (99.08\%)& & \\
\bottomrule
\end{tabularx}
\end{table}

\paragraph{Behavioral assessment criteria for HedWig}
The health checker sends $1911$ randomly selected emails from one account to another. We observe the following metrics: $1910$ out of $1911$ emails are successfully sent, fetched and validated. The single failed case is a validation error. Thus the percentage of validation errors VF=$1/1911=0.05\%$. There is no sending error, fetching error, or server crash found during the period, which means SF=FF=SC=$0\%$. These metrics form the behavioral assessment criteria that we use as baseline for each chaos engineering experiment.

\begin{table}[tb]
\centering
\scriptsize
\caption{Chaos Engineering Experiment Results on HedWig}\label{tab:resultsOfHedwig}
\begin{tabularx}{\columnwidth}{lrRXXXXXr}
\toprule
Target \& Error& F. Rate& Inj.& \multicolumn{6}{l}{Behavioral Assessment Criteria} \\
& & & SF& FF& VF& SC& CO& \\
\midrule
accept:EAGAIN& 0.4& 6& 40\%& 20\%& -& -& F& \colorbox{orange}{-}\\
accept:RANDOM& 0.5& 2& 99\%& 1\%& -& -& T& \colorbox{red}{!}\\
\hline
epoll\_ctl:ENOENT& 0.6& 5& -& -& -& -& F& \colorbox{green}{\makebox[0.3em]{√}}\\
epoll\_ctl:RANDOM& 0.5& 6& -& -& -& -& F& \colorbox{green}{\makebox[0.3em]{√}}\\
\hline
futex:EAGAIN& 0.05& 80& -& -& -& 33\%& F& \colorbox{red}{!}\\
futex:ETIMEDOUT& 0.588& 1.28K& -& -& -& 14\%& F& \colorbox{red}{!}\\
futex:RANDOM& 0.5& 886& -& -& -& -& F& \colorbox{green}{\makebox[0.3em]{√}}\\
\hline
read:EAGAIN& 0.24& 43& -& 83\%& -& -& F& \colorbox{orange}{-}\\
read:ECONNRESET& 0.05& 11& -& 57\%& 43\%& -& T& \colorbox{red}{!}\\
read:RANDOM& 0.5& 58& -& 67\%& -& 33\%& F& \colorbox{red}{!}\\
\hline
recvfrom:ECONNRESET& 0.05& 52& 75\%& 25\%& -& -& T& \colorbox{red}{!}\\
recvfrom:RANDOM& 0.5& 90& 99\%& 1\%& -& -& T& \colorbox{red}{!}\\
\hline
sendto:ECONNRESET& 0.05& 39& 12\%& 76\%& 12\%& -& F& \colorbox{orange}{-}\\
sendto:EPIPE& 0.05& 32& 30\%& 40\%& 30\%& -& F& \colorbox{orange}{-}\\
sendto:RANDOM& 0.5& 335& 97\%& 3\%& -& -& T& \colorbox{red}{!}\\
\bottomrule
\multicolumn{9}{p{8.5cm}}{Notes. If HedWig is resilient to the error, the result is marked as ``\colorbox{green}{\makebox[0.3em]{√}}''. If an injected error has an impact on the functionality only during error injection (SF, FF and VF related violations), the row is marked as ``\colorbox{orange}{-}''. If an injected error causes severe side effects like server crash (SC) or server state corruption (CO), the row is marked as ``\colorbox{red}{!}''}
\end{tabularx}
\end{table}

\paragraph{Chaos engineering experimental results}
The inputs for \autoref{algo:model-synthesis} are: the system call monitoring metrics $L=\autoref{tab:naturalErrorsOfHedwig}$, the upper boundary error rate $b=0.05$, a fixed error rate $r=0.05$ for sporadic errors, a threshold of the variance of error rates for fluctuating errors $t=0.001$, and a amplification factor $f=1.2$ for steady errors. In total, \phoebe synthesizes $13$ error models for chaos engineering experiments.

\autoref{tab:resultsOfHedwig} describes the results for those $9/13$ error models. We omit $4$ error models for which \phoebe did not inject any errors during the experiments because of the rarity of the corresponding system calls. To give a baseline, \autoref{tab:resultsOfHedwig} also presents the results for the random error injection approach which is described in \autoref{sec:experiment-protocol}.
Each row in the table records the results of one experiment, including the target system call, the injected error code, the applied error rate, the total number of injected errors, and HedWig's behavior under error injection. An error code \texttt{RANDOM} stands for the corresponding error injection experiment using the random injection strategy which is described in \autoref{sec:experiment-protocol}.

For example, the first row of \autoref{tab:resultsOfHedwig} is the result for a realistic error injection model based on the natural invocation error \texttt{accept:EAGAIN} (row 1 in \autoref{tab:naturalErrorsOfHedwig}) which meets case 2 in \autoref{sec:synthesizer}. During the experiment with error rate $0.4$, \phoebe injects $6$ errors in total into the system calls. $40.0\%$ of emails were successfully sent, fetched and verified. However, these errors caused sending errors for $40.0\%$ of emails and fetching errors for $20\%$ of emails. There was no validation error or server crash detected. The post inspection for this experiment passed, which means that the injected errors do not impact the server after the injection has been stopped. Consequently, failing invocations to \texttt{accept} do not cause a negative long-term impact on HedWig's functionalities.

\text{Recap.} The results of these chaos engineering experiments bring insights about how HedWig server behaves under different operating system perturbations. Based on the impact of a system call invocation error on Hedwig, the experiment results are categorized into three types: 1) an injected system call invocation error does not violate any behavior assessment criterion, is marked as ``\colorbox{green}{\makebox[0.3em]{√}}'' in the table, 2) an injected error has an impact on the functionality only during error injection (SF, FF and VF related violations), this is marked as ``\colorbox{orange}{-}'', and 3) an injected error causes severe side effects like server crash (SC) or server state corruption (CO), this is marked as ``\colorbox{red}{!}''. For example, \texttt{futex} related errors may lead the server to crash. Failing invocations to \texttt{read} and \texttt{recvfrom} can corrupt the application's running state and have a long-existing impact after an error happens. These categories of system call invocation errors are helpful to guide developers to design specific error handling mechanisms with respect to system call invocation errors.

\paragraph{Comparison against random injection}
Now we compare \phoebe's realistic error injection against random error injection. Recall \autoref{tab:resultsOfHedwig} shows the findings of both strategies, and indeed they yield different results. 
For example, the second row in the table gives the result of the random injection strategy for system call \texttt{accept}. For this call the sending and fetching failure  percentages are different from injection of realistic error code \texttt{EAGAIN} (the first row). Also, as shown in column SC, the random injection approach also causes a state corruption.: HedWig fails to operate normally even after the random injection stops. While a state corruption or a crash needs attention from developers, here this is due to an error unlikely to happen.

By analyzing those results, we identify two main reasons why \phoebe can considered better compared to  a random approach. 
1) The random approach may cause false alarms for developers. If a crash only happens because of some randomly generated errors which cannot occur in production, the alarm does not provide actionable feedback for the developer to improve the application and can be considered as noise.
2) The random approach may miss interesting corner cases because of lack of time. Error injection experiments for \texttt{futex} (row 5 to 7) are good examples here. When a realistic error code \texttt{EAGAIN} or \texttt{ETIMEDOUT} is injected by \phoebe, server crashes are observed; indicating that HedWig is not resilient to these error codes. However, the random injection strategy misses the injection of both these error codes and generates some other unrealistic codes instead. When performing fault injection under fixed time constraints, \phoebe successfully finds the interesting cases before random injection.

\subsection{Experiment on TTorrent}

\subsubsection{Experimental Settings for TTorrent}
We use version 2.0 of TTorrent for the experiments. As TTorrent uses BitTorrent protocol to download files from the Internet, its workload during file downloading can be considered as a production-like workload. To make the workload more various, TTorrent randomly downloads ubuntu-18.04.4, ubuntu-19.10, or ubuntu-20.04 using different torrent files for each experiment. According to the network condition and the experiment virtual machine's power, the average time of downloading one of the iso files in the data set is about $30$ seconds. For each experiment, the orchestrator adds a $150$ seconds time out for the TTorrent process. If TTorrent is still running but the file is not downloaded after $150$ seconds, the orchestrator kills the TTorrent process and begins the next round of download.

We run TTorrent to randomly download different Ubuntu distributions for $6$ hours without injecting errors, to determine its behavioral assessment criteria and observe the natural occurrences of system call invocation errors. Our key metric for the behavioral assessment criteria is the percentage of executions that successfully download files. When system call invocation errors are injected, TTorrent might behave in the following ways:
1) ST (stalled) TTorrent fails to download the file in a limited time, which is considered as stalled. The experiment orchestrator kills the TTorrent process and starts another round of experiment.
2) VF (validation error): TTorrent downloads the file, but the file's checksum is incorrect. This means a data corruption happens during the downloading process.
3) CR (crash): TTorrent directly crashes when an error is injected.
Besides these three kinds of abnormal behavior, it is also possible that TTorrent still successfully downloads the target file, with a correct md5 checksum. This means TTorrent's error handling mechanisms can handle the errors that are injected.

Since TTorrent is a client-side application, each chaos engineering experiment with TTorrent initializes a new process. There is no need to add post inspections for each experiment.

\subsubsection{Experimental Results}

\paragraph{Observation of natural system call errors}
The workload is a $6$-hour ($369$ separate executions in total) run of TTorrent to download different versions of Ubuntu distributions.

\phoebe's natural error monitor collects $16$ different system calls that naturally fail. They are summarized in~\autoref{tab:naturalErrorsOfTTorrent}. Each row in the table records one system call invocation error, including its system call name, error code, the total number of errors, the minimum (5th percentile), mean, maximum (95th percentile) value of the error rate per monitoring interval (15 seconds), and its corresponding cases for error model synthesis described in \autoref{sec:synthesizer}. In order to present the ratio of natural failures in each type of system call, \autoref{tab:naturalErrorsOfTTorrent} also records the number of successful system call invocations, which is denoted as \texttt{syscall:SUCCESS}.

For example, we observe $2.54K$ \texttt{access} system call invocation errors with an error code \texttt{ENOENT}, over the $369$ rounds of execution. This type of error takes $58\%$ of all invocations to system call \texttt{access}. The minimum, mean and maximum error rate per monitoring interval are respectively $0.57143$, $0.60784$ and $0.75000$. The variance of error rate during this observation time is $0.00584$. This indicates that the error rate varies a lot in different monitoring intervals: in some monitoring intervals, $57\%$ of the invocations to \texttt{access} fail with \texttt{ENOENT}, in some other monitoring intervals, $75\%$ of the invocations to \texttt{access} fail with such an error code.

\begin{table}[!tb]
\centering
\scriptsize
\caption{The Observed Natural Errors of TTorrent When Downloading a Ubuntu Distribution ISO File}\label{tab:naturalErrorsOfTTorrent}
\begin{tabularx}{\columnwidth}{lrrR}
\toprule
\textbf{Syscall \& Err.}& \textbf{Count}& \textbf{Error Rate (min, mean, max, var.)}& \textbf{Case} \\
& & \textbf{per 15 sec period}& \\
\midrule
access:ENO.& 2.54K (58\%)& 0.57143, 0.60784, 0.75000, 0.00584& 2\\
access:SUC.& 1.81K (42\%)& & \\
\hline
connect:EIN.& 7.17K (86\%)& 0.50000, 0.79292, 1.00000, 0.06331& 2\\
connect:ENO.& 676 (8\%)& 0.07143, 0.80854, 1.00000, 0.10869& 2\\
connect:SUC.& 517 (6\%)& & \\
\hline
epoll\_ctl:ENO.& 1.75K (<0.01\%)& 0.00002, 0.00608, 0.00187, 0.00558& 1\\
epoll\_ctl:SUC.& 9.14M (100\%)& & \\
\hline
epoll\_wait:EIN.& 46 (<0.01\%)& 0.50000, 0.73913, 1.00000, 0.06238& 2\\
epoll\_wait:SUC.& 10.7M (100\%)& & \\
\hline
futex:EAG.& 123K (<0.01\%)& 0.00112, 0.00638, 0.01546, 0.00003& 1\\
futex:ETI.& 452K (2\%)& 0.00455, 0.26513, 0.49319, 0.03666& 2\\
futex:SUC.& 24.6M (98\%)& & \\
\hline
gets.name:ENOT.& 169 (1\%)& 0.00969, 0.28275, 0.33333, 0.01302& 2\\
gets.name:SUC.& 12.9K (99\%)& & \\
\hline
lstat:ENO.& 507 (2\%)& 0.02256, 0.02342, 0.02381, <1e-5& 1\\
lstat:SUC.& 22.1K (98\%)& & \\
\hline
openat:EEX.& 169 (1\%)& 0.00420, 0.00596, 0.00658, <1e-5& 1\\
openat:ENO.& 7.92K (20\%)& 0.08696, 0.22214, 0.35706, 0.01536& 2\\
openat:SUC.& 31K (79\%)& & \\
\hline
read:EAG.& 335 (<0.01\%)& 0.00001, 0.00003, 0.00009, <1e-5& 1\\
read:ECO.& 186 (<0.01\%)& 0.00001, 0.00003, 0.00012, <1e-5& 1\\
read:SUC.& 21.3M (100\%)& & \\
\hline
stat:ENO.& 11.5K (36\%)& 0.11429, 0.36465, 0.57143, 0.02938& 2\\
stat:SUC.& 20.7K (64\%)& & \\
\hline
unlink:ENO.& 246 (59\%)& 0.66667, 0.78320, 1.00000, 0.02526& 2\\
unlink:SUC.& 169 (41\%)& & \\
\hline
write:ECO.& 29 (<0.01\%)& 0.00001, 0.00003, 0.00006, <1e-5& 1\\
write:SUC.& 9.09M (100\%)& & \\
\bottomrule
\end{tabularx}
\end{table}

\paragraph{Behavioral assessment criteria for TTorrent}
During the observation of natural errors, it shows that $358$ out of $369$ executions successfully download the target file with a correct MD5 checksum. The other $11$ executions all lead to a stalled state, ST=$3.0\%$. There is no validation error or crash detected, VF=CR=$0\%$. These metrics are used as a control group to detect abnormal behavior during chaos engineering experiments.

\begin{table}[tb]
\centering
\scriptsize
\caption{Chaos Engineering Experiment Results on TTorrent}\label{tab:resultsOfTTorrent}
\begin{tabularx}{\columnwidth}{lrrXXXr}
\toprule
Target \& Error& F. Rate& Inj.& \multicolumn{4}{l}{Behavioral Assessment Criteria} \\
& & & ST& VF& CR& \\
\midrule
access:ENOENT& 0.9& 71& -& -& 100\%& \colorbox{red}{!}\\
access:RANDOM& 0.5& 27& 8\%& -& 75\%& \colorbox{red}{!}\\
\hline
connect:EINTR& 1.0& 72& 100\%& -& -& \colorbox{orange}{-}\\
connect:ENOENT& 1.0& 72& 100\%& -& -& \colorbox{orange}{-}\\
connect:RANDOM& 0.5& 7& 25\%& -& -& \colorbox{orange}{-}\\
\hline
epoll\_ctl:ENOENT& 0.05& 29.2K& -& -& -& \colorbox{green}{\makebox[0.3em]{√}}\\
epoll\_ctl:RANDOM& 0.5& 3& 100\%& -& -& \colorbox{orange}{-}\\
\hline
epoll\_wait:EINTR& 1.0& 5.33M& -& -& -& \colorbox{green}{\makebox[0.3em]{√}}\\
epoll\_wait:RANDOM& 0.5& 15.1K& -& -& -& \colorbox{green}{\makebox[0.3em]{√}}\\
\hline
futex:EAGAIN& 0.05& 97& -& -& 95\%& \colorbox{red}{!}\\
futex:ETIMEDOUT& 0.493& 69& -& -& 100\%& \colorbox{red}{!}\\
futex:RANDOM& 0.5& 5.93K& -& -& -& \colorbox{green}{\makebox[0.3em]{√}}\\
\hline
getsockname:ENOTSOCK& 0.333& 59& 50\%& -& -& \colorbox{orange}{-}\\
getsockname:RANDOM& 0.5& 40& 67\%& -& -& \colorbox{orange}{-}\\
\hline
openat:EEXIST& 0.05& 48& 8\%& -& 75\%& \colorbox{red}{!}\\
openat:ENOENT& 0.428& 121& -& -& 100\%& \colorbox{red}{!}\\
openat:RANDOM& 0.5& 94& -& -& 100\%& \colorbox{red}{!}\\
\hline
read:EAGAIN& 0.05& 198& -& -& 100\%& \colorbox{red}{!}\\
read:ECONNRESET& 0.05& 226& -& -& 100\%& \colorbox{red}{!}\\
read:RANDOM& 0.5& 187& -& -& 100\%& \colorbox{red}{!}\\
\hline
unlink:ENOENT& 1.0& 8& -& -& -& \colorbox{green}{\makebox[0.3em]{√}}\\
unlink:RANDOM& 0.5& 3& 20\%& -& -& \colorbox{orange}{-}\\
\hline
write:ECONNRESET& 0.05& 131& 100\%& -& -& \colorbox{orange}{-}\\
write:RANDOM& 0.5& 1.04K& 100\%& -& -& \colorbox{orange}{-}\\
\bottomrule
\multicolumn{7}{p{8.5cm}}{Notes. If TTorrent always successfully downloads the file with a correct checksum under error injection, the row is marked as ``\colorbox{green}{\makebox[0.3em]{√}}''. If TTorrent gets stalled when an error is injected, the row is marked as ``\colorbox{orange}{-}''. If TTorrent reports that the file is downloaded but the checksum of the file is incorrect, or TTorrent immediately crashes after injecting an error, the row is marked as ``\colorbox{red}{!}''}
\end{tabularx}
\end{table}

\paragraph{Chaos engineering experiment results}
From those natural errors, \phoebe synthesizes $16$ realistic error injection models for chaos engineering experiments with the following configuration. The upper boundary error rate $b = 0.05$, the fixed error rate for rare system calls is $0.05$, the threshold of the variance $t=0.001$, and the amplification factor $f = 1.2$. The generated error models and the corresponding experiment results are presented in \autoref{tab:resultsOfTTorrent}. Similar to \autoref{tab:resultsOfHedwig}, the rows where the injection count is zero are omitted as well. The experiment results are categorized into $3$ types: 1) TTorrent always successfully downloads the file with a correct checksum under error injection, which is marked as ``\colorbox{green}{\makebox[0.3em]{√}}'' in the table, 2) TTorrent gets stalled when an error is injected (ST>0, and VF=CR=0), which is marked as ``\colorbox{orange}{-}'', 3) TTorrent reports that the file is downloaded but the checksum of the file is incorrect (VF>0), or TTorrent immediately crashes after injecting an error (CR>0), which is marked as ``\colorbox{red}{!}''.

\text{Recap.}
\phoebe finds that TTorrent has different levels of reliability depending on the system call invocation errors. For example, TTorrent is fully resilient against errors of types \texttt{epoll\_ctl:ENOENT}, \texttt{epoll\_wait:EINTR} and \texttt{unlink:ENOENT}, as shown in rows $6$, $8$ and $21$. In other words, those errors had no negative impact on the behavior assessment criteria. In those case, TTorrent was still able to download the file under despite error injection. However, \phoebe also shows that TTorrent is particularly sensitive to invocation errors related to \texttt{access}, \texttt{futex}, \texttt{openat} and \texttt{read}, where error injection make TTorrent crash instantly. 

\paragraph{Comparison against random injection}
\autoref{tab:resultsOfTTorrent} enables us to compare error injection with \phoebe and with a random error injection strategy. Overall, the findings are similar to those for HedWig. 
First, the random injection strategy produces some false alarms for developers, such as those for system call \texttt{epoll\_ctl} and \texttt{unlink}. 
Second, the random injection strategy misses to provide valuable insights about some specific errors, such as system call \texttt{futex} because of lack time to visit the whole injection space \cite{durieux:hal-01624988}. 
For example, let us consider the injection of errors into invocations of system call \texttt{futex} (row 10-12 in the table), Phoebe tells the developers that a small proportion of errors with EAGAIN yields state corruption, which is a concrete target for engineers. On the contrary, random unrealistic error codes appear to be acceptable.

\section{Discussion}
\label{sec:discussion}

\subsection{Runtime Overhead of \phoebe}

\emph{Overhead Evaluation Protocol}
Monitoring and injecting system call invocation errors may impact application performance. We now measure and discuss the runtime overhead of \phoebe. Firstly, we keep an application running in production for a certain amount of time without \phoebe attached, and we record performance-related metrics. Secondly, the application is executed for the same duration with \phoebe's monitor attached. The same performance metrics are captured. Finally, the performance difference is analyzed to determine if \phoebe has an acceptable runtime overhead.

We collect both generic and application-specific performance metrics. The generic metrics are: 1) heap memory usage in the JVM collected with Glowroot, 2) CPU load. Since HedWig depends on a MySQL database, the JDBC transactions are also picked up as an application-specific performance metric. For TTorrent, we measure the average download time. 


\emph{Overhead Evaluation Results}
\autoref{tab:overhead} shows the runtime overhead of \phoebe's natural error monitor on HedWig and TTorrent. For example, the first group of rows in the table shows the $5$ metrics captured for HedWig: 1) the heap memory usage, 2) the CPU load, 3) the memory usage per database transaction, 4) the CPU time per transaction and 5) the JDBC query time on average.
These $5$ metrics respectively increase by $5.0\%$, $5.9\%$, $0.4\%$, $7.4\%$, and $8.0\%$ with system call monitoring, which is considered as acceptable. The same conclusion applies to TTorrent, where the maximum overhead is 4.9\%.


As a summary, the runtime overhead cost by \phoebe is comparable to other monitoring tools for production usage like Glowroot \cite{glowroot-overhead} and SWAT \cite{Hauswirth:SWAT}.

\begin{table}[!tb]
\centering
\scriptsize
\caption{The Runtime Overhead of \phoebe's Natural Error Monitor during Executing HedWig (H.) and TTorrent (T.)}\label{tab:overhead}
\begin{tabularx}{\columnwidth}{llrrr}
\toprule
\textbf{App.}& \textbf{Metric}& \textbf{Normal Run}& \textbf{Monitor On}& \textbf{Overh.}\\
\midrule
\multirow{5}{*}{H.}& Heap Memory Usage& 104.6MB& 109.8MB& 5.0\%\\
& CPU Load& 1.7\%& 1.8\%& 5.9\%\\
& Memory Usage per Transaction& 837.5 KB& 840.9 KB& 0.4\%\\
& CPU Time per Transaction& 20.3 ms& 21.8 ms& 7.4\%\\
& JDBC Query Average Time& 0.351 ms& 0.379 ms& 8.0\%\\
\midrule
\multirow{3}{*}{T.}& Heap Memory Usage& 150.2MB& 156.9MB& 4.5\%\\
& CPU Load& 31.9\%& 32.6\%& 2.2\%\\
& Average Downloading Time& 36.8s& 38.6s& 4.9\%\\
\bottomrule
\end{tabularx}
\end{table}

\subsection{Applicability to Other Software Stacks}
As \phoebe's natural error monitor and system call error injector are implemented using the eBPF feature of the Linux kernel, \phoebe's capability of error model synthesis and error injection is applicable to all Linux executables (as long as the kernel is built with eBPF activated). If the goal is to conduct realistic error injection experiments and observe their impact on general metrics such as the number of crashes, \phoebe is fully appliable to other software stacks, different from Java.
However, in order to get fine-grained observability under error injection experiments, \phoebe requires an application behavior monitor which is  specifically implemented for Java applications. This is necessary because \phoebe collects application-specific metrics that are based on the JVM.

\subsection{Threats to Validity}

A critical bug in \phoebe that impacts the trustfulness of our measurements would impact internal validity. Since our code is open-source, future work and researchers in this domain are able to verify it.

The behavior assessment criteria are essential for analyzing the application behavior under fault injection. If we have missed an application-specific behavior assessment criterion, this would impact construct validity. However, we are confident that this is not the case since we understand the domain of email communication and file downloading.

Finally, \phoebe focuses on applications running on top of the Java Virtual Machine because application behavior monitoring requires language- and technology-specific features. We notice that the JVM itself may 1) either create natural system call errors 2) or remediate some errors directly (i.e. we are not observing the effectiveness of application-specific reliability). Future work may explore the interplay between the JVM and the application error-handling mechanisms with respect to system call invocation errors.

\section{Related Work}\label{sec:related-work}

In this section, we discuss work related to fault injection and observability.

\begin{table*}[tb]
\centering
\caption{Comparison of \phoebe with the closest related work}\label{tab:related-work}
\begin{tabularx}{\textwidth}{lRRRR}
\toprule
\textbf{Dimension}& \textbf{Ballista \cite{Koopman:Ballista}}& \textbf{Syzkaller \cite{vyukov2019syzkaller}}& \textbf{ChaosMachine \cite{Zhang:chaosmachine}}& \textbf{\phoebebf} \\
\midrule
Fault Injection Target& Operating systems& Operating systems& Java Applications& \textbf{Java Applications}\\
Perturbation model& All possible invalid inputs for syscalls& All possible invalid inputs for syscalls& Manually configured Java exceptions& \textbf{Auto-generated realistic syscall invocation errors}\\
Running Env.& Testing Env.& Testing Env.& JVM in production& \textbf{JVM in production}\\
Performance overhead& Not evaluated& Not evaluated& Low& \textbf{Low}\\
\bottomrule
\end{tabularx}
\end{table*}

\subsection{Fault Injection}
Fault injection techniques aim at evaluating the error handling mechanisms of a software system \cite{Natella:Assessing_Dependability_with_Software_Fault_Injection_A_Survey,Ziade:A_Survey_on_Fault_Injection_Techniques,Silva:A_view_on_the_past_and_future_of_fault_injection_2013}. Fault injection research has heavily focused on hardware errors \cite{Karlsson:heavy-ion-radiation,madeira1994rifle}. Another line of research work concerns the injection of high-level software faults \cite{Hyosoon:SFIDA_a_software_implemented_fault_injection_tool_for_distributed_dependable_applications,Kao:A_fault_injection_and_monitoring_environment_for_tracing_the_UNIX_system_behavior_under_faults,Kouwe:HSFI_Accurate_Fault_Injection_Scalable_to_Large_Code_Bases:DSN}.

Regarding hardware-related fault injection such as bit flips, Kanawati et al. \cite{kanawati1992ferrari} proposed FERRARI, a software system that emulates hardware faults. Han et al. \cite{han1993doctor} designed DOCTOR, which focuses on injecting hardware errors and network errors. Carreira et al. \cite{Carreira:TSE1998:Xception} presented Xception, a fault injection and monitoring environment that emulates processor failures. Wei et al. \cite{Wei:Quantifying_the_Accuracy_of_High-Level_Fault_Injection_Techniques_for_Hardware_Faults:DSN} quantitatively evaluated the accuracy of intermediate code level fault injection with respect to assembly level fault injection for hardware-related errors. Fang et al. \cite{Fang:ISPASS2014:gpu-qin} presented GPU-Qin, which injects bit-flip faults into a GPU's register to evaluate end-to-end reliability properties of application kernels running on GPUs. Guan et al. \cite{Guan:IPDPS2014:F-SEFI} proposed F-SEFI, a tool that leverages the QEMU emulator to introduce bit-flip errors for a specific sub-function in an application.

Regarding high-level software error injection such as operating system faults, Lee et al. \cite{Hyosoon:SFIDA_a_software_implemented_fault_injection_tool_for_distributed_dependable_applications} presented SFIDA, which is used to evaluate the resilience of distributed applications on the Linux platform. Kao et al. \cite{Kao:A_fault_injection_and_monitoring_environment_for_tracing_the_UNIX_system_behavior_under_faults} invented ``FINE'', a fault injection and monitoring tool to inject both hardware-induced software errors and software faults.
Bagchi et al. \cite{Bagchi:DSOM2001:Dependency_Analysis_in_Distributed_Systems_using_Fault_Injection} applied fault injection as the perturbation tool for dynamic dependency discovery and problem determination in an e-commerce environment.
Duraes et al. \cite{Duraes:TSE2006:emulation-of-software-faults} presented G-SWFIT (Generic Software Fault Injection Technique) that injects faults directly in the target executable code.
Kouwe and Tanenbaum \cite{Kouwe:HSFI_Accurate_Fault_Injection_Scalable_to_Large_Code_Bases:DSN} presented HSFI, which takes execution context information into consideration for efficient fault injection decisions.
Cotroneo et al. \cite{cotroneo2020profipy} proposed ProFIPy, a tool that is programmable to specify different fault models using a domain-specific language for fault injection in Python.
Zhang et al. \cite{Zhang:ISSRE2019:TripleAgent} designed TripleAgent, a tool that combines automated monitoring, exception injection, and resilience improvement for Java applications.

More related to our work, there are previous fault injection approaches related to system call invocation errors. Koopman and DeVale \cite{Koopman:Ballista} proposed Ballista, a testing system that generates invalid inputs for system call invocations in order to evaluate the exception handling effectiveness of POSIX operating systems. Vyukov \cite{vyukov2019syzkaller} designed syzkaller, a tool that fuzzes system call invocation inputs in order to detect kernel bugs. Amarnath et al. \cite{amarnath2018fault} designed a QEMU-based fault injection framework to evaluate the dependability of system calls with respect to bit flip errors. Simonsson et al. \cite{simonsson2019observability} presented ChaosOrca, a chaos engineering system for dockerized applications.

Regarding realism of fault injection, Natella et al. \cite{Natella:TSE2013:OnFaultRepresentativeness} evaluated the representativeness of faults injected by G-SWFIT with the help of test suites. They proposed a new approach to refine the fault model by filtering out non-representative faults in the searching space. The main difference between their work and \phoebe is that \phoebe improves the realism by analyzing naturally happening errors in production. Natella et al.'s technique is entirely different, since they conduct offline fault injection experiments and then remove the non-representative ones.

In this past related work, the error models are either randomly generated or predefined by developers. On the contrary, \phoebe exclusively focuses realistic error models synthesized from errors that naturally happen. To highlight the novelty of \phoebe, \autoref{tab:related-work} makes a comparison between \phoebe and the closest related work. There are four different dimensions under comparison: 1) the fault injection software targets under evaluation, 2) the perturbation model, 3) the running environment, and 4) the performance overhead. As seen in this table, \phoebe is novel and unique in what it does: system call error injection in production.

\subsection{Chaos Engineering}

Chaos engineering can be defined as doing high-level fault injection on the production system directly \cite{Basiri2017:chaos-engineering-book}. Netflix's ChaosMonkey \cite{Chang:Chaos_Monkey_Increasing_SDN_Reliability_Through_Systematic_Network_Destruction} randomly shuts down servers in production in order to verify the whole system's reliability against a server crash. Then this methodology has been extended with other kinds of errors such as the OS level and the network errors \cite{The_Netflix_Simian_Army,Heather:Inside_Azure_Search_Chaos_Engineering}. There is also application-level chaos engineering research: Sheridan et al. \cite{Sheridan:DICE_Fault_Injection_Tool} presented a fault injection tool for cloud applications, where faults are resource stress or service outage; Zhang et al. \cite{Zhang:chaosmachine} devised ChaosMachine, a chaos engineering system that analyzes a Java application's exception-handling capabilities in production.

To the best of our knowledge, none of the existing chaos engineering approaches synthesize realistic error models based on naturally happening errors like \phoebe.

\subsection{Observability}

Monitoring techniques are most widely researched in the area of observability. Grobmann and Klug \cite{GrobmannMarcel2017MCSa} proposed ``PyMon'', a framework that monitors different computing architectures with a small footprint. Povedano-Molina et al. \cite{Povedano-Molina:DARGOS} designed DARGOS, a distributed architecture for resource management and monitoring in cloud computing. Arora et al. \cite{Arora:2018:RWR:3238147.3238186} presented a system called Parikshan that duplicates traffic into a copy of the production container, enabling the use of heavier monitoring tools without impacting the performance in production. Chang et al. \cite{Chia-ChenChang2017AKMP} developed a Kubernetes-based monitoring platform for dynamic cloud resource provisioning. Enes et al. \cite{ENES2018420:BDWatchdog} proposed BDWatchdog, a solution for real-time analysis of big data frameworks and workloads that combines per-process resource monitoring and low-level profiling.

Another popular research direction in observability is tracing. Sigelman et al. \cite{Sigelman:Dapper} presented Google's Dapper, a tracing infrastructure with low overhead and application-level transparency. Kaldor et al. \cite{Kaldor:Canopy} presented Canopy, Facebook’s end-to-end performance tracing infrastructure that enables developers to analyze performance data in real-time. Mace et al. \cite{Mace:PivotTracing} presented Pivot Tracing, which implements a happened-before join operator to enhance dynamic instrumentation and causal tracing. Instead of analyzing network requests in a distributed system, Coppik et al. \cite{Coppik:TrEKer} proposed TrEKer, an approach that combines static and dynamic analyses to trace error propagation in OS kernels. Ren et al. \cite{Ren:RepTrace} invented RepTrace, a framework that traces system calls in order to analyze the root causes of unreproducible builds. The traces of system call invocations also helps anomaly detection. Liu et al. \cite{Liu:syscall_traces_for_anomaly_detection} proposed a feature extraction method named STP that transforms the system call sequences into frequency sequences of n-grams in a trace to detect abnormal behavior. Cotroneo et al. \cite{Cotroneo:TDSC2020:FIAnalytics} utilized tracing and machine learning techniques to cluster injected failures into different models for a distributed system.

Differently, \phoebe focuses on the observability of system call invocation errors, which is not in the scope of this related work. Furthermore, \phoebe synthesizes error injection models based on the monitoring observation. The combination of monitoring in production and fault injection is original.

\section{Conclusion}\label{sec:conclusion}
In this paper, we have presented \phoebe, a novel fault injection framework for reliability evaluation against system call invocation errors. The key novelty of \phoebe is that it synthesizes and injects realistic system call errors, meaning that the injected errors are based on focusing on errors that naturally happen.
By evaluating \phoebe's functionality and performance on two medium-sized real-world applications (email server, file transfer), we have shown that it is able 1) to detect reliability weaknesses 2) with low overhead. In the future, we will study the relationship between low-level system call invocation errors and high-level Java exceptions. This would developers to identify concrete locations in thee application source code to fix the reliability weaknesses detected by \phoebe.

\section*{Acknowledgements}
This work was partially supported by the Wallenberg AI, Autonomous Systems and Software Program (WASP) funded by the Knut and Alice Wallenberg Foundation.

\bibliographystyle{plain}
\balance
\bibliography{references}

\end{document}